\documentclass[12pt]{article}
\usepackage{amssymb,amsmath,latexsym,graphicx, bm, mathrsfs, hyperref}
\usepackage{graphics,graphicx}
\usepackage[dvipsnames]{xcolor}
\begin{document}

\title{Jacobi-Maupertuis Randers-Finsler metric for curved spaces
\\ {and the  gravitational magnetoelectric effect}\author{Sumanto Chanda$ ^1$, G. W. Gibbons$ ^2$, Partha Guha$ ^3$, \\ Paolo Maraner$ ^4$, and Marcus C. Werner$ ^5$}}

\maketitle

\begin{minipage}{0.3\textwidth}
\begin{flushleft}
\textit{$ ^1$ International Center \\ for Theoretical Sciences}\\ 
\textit{\small No. 151, Shivakote, Hesaraghatta Hobli}, \\ 
\textit{\small Bengaluru (North, Karnataka) 560089, INDIA.} \\
\texttt{\small sumanto.chanda@icts.res.in}
\end{flushleft}
\end{minipage}
\begin{minipage}{0.3\textwidth}
\begin{center}
\textit{$ ^2$  D.A.M.T.P., \\ University of  Cambridge} \\
\textit{\small Wilberforce Road, \\ Cambridge CB3 0WA,  U.K.} \\
\texttt{\small G.W.Gibbons@damtp.cam.ac.uk \\ gwg1@cam.ac.uk}
\end{center}
\end{minipage}
\begin{minipage}{0.3\textwidth}
\begin{flushright} \large
\textit{\small $ ^3$ S.N. Bose National Centre for Basic Sciences} \\
\textit{\small JD Block, Sector-3, Salt Lake, Calcutta-700098, INDIA.} \\
\texttt{\small partha@bose.res.in}
\end{flushright}
\end{minipage} \\ 

\begin{center}
\begin{minipage}{0.42\textwidth}
\begin{flushleft}
\textit{$ ^4$ {School of Economics and Management},\\ 
Free University of Bozen-Bolzano,} \\
\textit{Universit\"atsplatz 1 - Piazza Universit\`a 1,} \\ 
\textit{\small  39100 Bozen-Bolzano, ITALY.} \\
\texttt{\small pmaraner@unibz.it}
\end{flushleft}
\end{minipage}
\begin{minipage}{0.42\textwidth}
\begin{flushright} \large
\textit{\small $ ^5$ YITP Center for Gravitational Physics}\\ 
\textit{\small Kyoto University,} \\ 
\textit{\small Kitashirakawa Oiwakecho Sakyoku,} \\ 
\textit{\small Kyoto 606-8502, JAPAN.} \\
\texttt{\small werner@yukawa.kyoto-u.ac.jp}
\end{flushright}
\end{minipage}
\end{center}

\begin{abstract} 
In this paper we return to the subject of Jacobi metrics for
timelike and null geodsics in stationary spactimes, correcting some previous misconceptions.
We show  that  not  only null geodesics, but also timelike geodesics are governed by a Jacobi-Maupertuis type
variational  principle and a Randers-Finsler metric for which we give  explicit formulae.
The cases of the Taub-NUT and Kerr spacetimes are discussed in detail. Finally we show
how our  Jacobi-Maupertuis Randers-Finsler metric may be expressed
in terms of the effective medium describing the
behaviour of Maxwell's equations in the  curved spacetime. In particular,
we see in very concrete terms how the magnetolectric susceptibility
enters the  Jacobi-Maupertuis-Randers-Finsler function.
\end{abstract}

\section{Introduction}

In this paper we consider the motion of a neutral particle moving in a
stationary spacetime. Because the metric is stationary there is a
conserved energy and it is natural, following the  time honoured
procedure of Maupertuis and Jacobi, to convert the problem to a
variational problem at fixed energy. If the metric is static this is
straight forward and one finds that the motion may be obtained as the
the geodesics of a Riemannian metric which for massive particles
depends on the conserved energy \cite{gwg}. For massless particles the
metric is independent of the energy and is often referred to as the
optical or sometimes the Fermat metric.  For stationary metrics the
cross term between the space and time components of the spacetime
metric renders the situation more complicated.

For massless particles the motion may be obtained as the geodesic of a
metric which is, however, not Riemannian but rather a Finsler metric
of so-called Randers type \cite{Gibbons:2008zi}.  Randers-Finsler
metrics arose originally in an attempt to  provide a unified
  description of gravitation and electromagnetism.  As a first step
  Randers started by considering a charged particle moving in curved
  spacetime metric and an electromagnetic field\cite{Randers:1941gge}.
  Subsequently the particle trajectories were recognised as the
  geodesics of a particular example of a Finsler metric.  In
\cite{gwg} it was stated that ``the motion of massive particles is
governed by neither a Riemannian nor a Finslerian metric''. As pointed
out in \cite{paolo} this statement is not correct. In fact, the
motions of massive neutral particles are the geodesics of an energy
dependent Randers-Finsler metric. In this paper we elaborate on the
considerations of \cite{paolo} and  the light they throw on the
unsuccessful  attempts  of  \cite{jmp}. 

As pointed out above, the necessity of passing to a Randers-Finsler metric arises from the cross terms in the metric. This is perhaps
not too surprising since the cross terms also give rise to the phenomenon sometimes referred to as the  ``rotation of inertial frames'' or, more graphically, ``gravito-magnetism''. In  \cite{Randers:1941gge}
the Maxwell field was treated as a fixed background. However it is also of interest to ask how a general time dependent Maxwell field responds to a stationary background.  One approach to this question is to write out Maxwell's equations in a local coordinate system defining electric and magnetic fields and electric and magnetic inductions.  The former are related to the latter by linear constitutive relations where the susceptibilities depend upon the components of the spacetime metric and are, in general, position dependent\cite{FrolovShoom,Castillo}.  In a stationary metric gravito-magnetism expresses itself as  gravitational magnetoelectric effect\cite{GibbonsWerner} in which the constitutive relations mix electric and magnetic fields. We are able to show how the susceptibilities enter the Jacobi-Randers-Finsler metric reflecting this phenomenon.

Furthermore, we know that the existence of a conserved quantity 
implies the reduction of a degree of freedom from 
the overall motion. Since the formulation of the Jacobi-Maupertuis metric is centred around the 
availability of a conserved quantity, it is essentially a reduction of the dimension of the moduli  space of geodesics. 
When the conserved quantity is associated with a cyclical co-ordinate, we are reducing the dimensions of the configuration space of the geodesic by 
that cyclical co-ordinate, thus simplifying the mechanical problem.

The plan of the paper is as follows. Section 2 introduces constraints among the spatial momenta and the energy giving rise to the difficulties experienced in \cite{gwg,jmp}. In section 3 we treat the simpler electromagnetic case in preparation for the full stationary case in section 4. The results obtained in section 4  agree with a remark in \cite{Perlick}, unkown to the authors at the time of writing \cite{gwg,jmp},
whose main purpose was to discuss relativistic brachistochrones.
  The  analogy to the motion of a charged particle moving in  electromagnetic field was noted  in \cite{Perlick} but not related
 to  Randers-Finsler geometry. In section 5 we give some explicit examples including the physically important case of the Kerr metric for a rotating black hole. Finally, in section 6 we relate the previous sections to the gravitational magnetoelectric effect.

\section{Constraint for momenta of a geodesic}

If the relativistic Lagrangian for a free particle moving in a stationary spacetime is given by:
\begin{equation} \label{stationary}
\mathcal L = - mc \sqrt{g_{\mu \nu} (\bm x)  \dot x^\mu \dot x^\nu}
  ={-} mc \sqrt{{c^2}g_{00} (\bm x) \dot t^2 + 2{c} g_{0i} (\bm x) \dot t \dot x^i  + g_{ij} (\bm x) \dot x^i \dot x^j},
\end{equation}
where the metric signature is $+,-,-,-$,
$\bm x=(x,y,z)$ specifies the spatial coordinates and the dot indicates differentiation with respect to an arbitrary parameter, then the canonical momenta are:
\begin{equation} \label{mom}
\begin{split}
p_\mu = \frac{\partial \mathcal L}{\partial \dot x^\mu} ={-}  mc \frac{g_{\mu \nu} (\bm x) \dot x^\nu}{\sqrt{g_{\alpha \beta} (\bm x)  \dot x^\alpha \dot x^\beta}}
\end{split} \qquad 
\left\{ \begin{split}
  p_0 = \frac{\partial \mathcal L}{{c}\partial \dot t} &= - mc
  \frac{{c} g_{00} (\bm x) \dot t +  g_{0i} (\bm x) \dot x^i}
  {\sqrt{g_{\alpha \beta} (\bm x)  \dot x^\alpha \dot x^\beta}}
  {=-\frac{\mathcal{E}}{c}} \\
  p_i = \frac{\partial \mathcal L}{\partial \dot x^i} &={-}  mc
  \frac{c g_{0i} (\bm x) \dot t + g_{ij} (\bm x) \dot x^j}{\sqrt{g_{\alpha \beta} (\bm x)  \dot x^\alpha \dot x^\beta}} 
\end{split} \right. .
\end{equation}
where $\mathcal{E} = p_i \dot{x}^i - \mathcal L$ is the relativistic energy of the system. Thus, the canonical momenta of a geodesic satisfy the constraint:
\begin{equation} \label{constraint}
g^{\mu \nu} (\bm x) p_\mu p_\nu = (m c)^2 g^{\mu \nu} (\bm x) \frac{\left( g_{\mu \alpha} (\bm x) \dot x^\alpha \right)}{\sqrt{g_{\rho \sigma} (\bm x)  \dot x^\rho \dot x^\sigma}} \frac{\left( g_{\nu \beta} (\bm x) \dot x^\beta \right)}{\sqrt{g_{\rho \sigma} (\bm x)  \dot x^\rho \dot x^\sigma}} = (m c)^2.
\end{equation}
The relativistic momenta are therefore constrained to the ``mass shell'' at all points of spacetime. On the other hand, for the Jacobi metric {$ds_{\!J}=p_i dx^i$}, one demands:
\begin{equation} \label{effact}
L_J  {\ \equiv \ } \dot{s}_{\!J} 
 = p_i \dot x^i = {-mc}\frac{g_{i \nu} \dot x^i \dot x^\nu }{\sqrt{g_{\alpha \beta} (\bm x) \dot x^\alpha \dot x^\beta}} \quad =\quad \sqrt{J_{ij} \dot x^i \dot x^j}.
\end{equation}
Thus, the momenta would be given by
\begin{equation} \label{jmom}
p_i = \frac{J_{ij} \dot x^j}{\sqrt{J_{ab} \dot x^a \dot x^b}}
\end{equation}
and the constraint equation for the momenta (\ref{jmom}) for the Jacobi metric would be:
\begin{equation} \label{jconstr}
J^{ij} (\bm x) p_i p_j = J^{ij} \frac{J_{ip} \dot x^p}{\sqrt{J_{ab} \dot x^a \dot x^b}} \frac{J_{iq} \dot x^q}{\sqrt{J_{ab} \dot x^a \dot x^b}} = \frac{\left(J^{ij} J_{ip} J_{iq} \right) \dot x^p \dot x^q}{J_{ab} \dot x^a \dot x^b} = 1.
\end{equation}
This rule has been applied to static spacetimes in \cite{gwg} (where $g_{0i} (\bm x) = 0 \ , \ g^{00} (\bm x) = \left( g_{00} (\bm x) \right)^{-1}$)
$$g^{00} (\bm x) p_0^2 + g^{ij} (\bm x) p_i p_j = (mc)^2 \quad \Rightarrow \quad \frac{g^{ij} (\bm x)}{(mc)^2 - g^{00} (\bm x) p_0^2} p_i p_j = \frac{g_{00} (\bm x) g^{ij} (\bm x)}{(mc)^2 g_{00} (\bm x) - p_0^2} p_i p_j = 1$$

$$ \Rightarrow \quad J^{ij} (\bm x) = \frac{g_{00} (\bm x)}{(mc)^2 g_{00} (\bm x) - p_0^2} g^{ij} (\bm x) \quad \Rightarrow \quad J_{ij} (\bm x) = \frac{(mc)^2 g_{00} (\bm x) - p_0^2}{g_{00} (\bm x)} g_{ij} (\bm x),$$
therefore
\begin{equation} \label{cnstrjm}
  {ds_{\!J}=\sqrt{\frac{(mc)^2g_{00} (\bm x) - p_0^2}{g_{00} (\bm x)} g_{ij} (\bm x)dx^idx^j}}
\end{equation}
and attempted in \cite{jmp}.
The constraint for momentum of a stationary spacetime will be:
$$g^{00} (\bm x) p_0^2 + 2 g^{0i} (\bm x) p_0 p_i + g^{ij} (\bm x) p_i p_j = (mc)^2,$$
from which it is evidently impossible to determine the constraint equation for the Jacobi metric {(\ref{effact})}
, compelling us to return to first principles. \smallskip

\section{Jacobi metric for a Lagrangian with magnetic fields}

In this section we shall proceed by considering a non-relativistic Lagrangian involving magnetic fields. Starting from the relativistic Lagrangian (\ref{stationary}), defining $g_{00}(\bm x)$ and parametrizing with respect to time
$$g_{00} (\bm x) = 1 + \frac{2 U (\bm x)}{mc^2} \quad , \quad \dot t = 1,$$ 
and taking non-relativistic approximation by expanding binomially up to first order as shown in \cite{cg}, we will have the non-relativistic Lagrangian involving magnetic fields:  
\[ \begin{split}
\mathcal L &= - mc \sqrt{g_{\mu \nu} (\bm x) \dot x^\mu \dot x^\nu} = - mc^2 \sqrt{1 + \frac2{m c^2} \left( \frac m2 g_{ij} (\bm x) \dot x^i \dot x^j + m c g_{0i} (\bm x) \dot x^i + U (\bm x) \right)} \\
&\approx - mc^2 \left[ 1 + \frac1{m c^2} \left( \frac m2 g_{ij} (\bm x) \dot x^i \dot x^j + m c g_{0i} (\bm x) \dot x^i + U (\bm x) \right) \right] \\
&= \left( - \frac m2 g_{ij} (\bm x) \dot x^i \dot x^j - m c g_{0i} (\bm x) \dot x^i - U (\bm x) \right) - mc^2 \qquad = \qquad L - mc^2.
\end{split} \]
Omitting the additive term $- mc^2$, and writing $G_{ij} (\bm x) = - g_{ij} (\bm x) \ , \ - m c g_{0i} (\bm x) = A_i (\bm x)$, we get:
\begin{equation} \label{maglag}
L = \frac m2 {G}_{ij} (\bm x) \dot x^i \dot x^j + A_i (\bm x) \dot x^i - U (\bm x)\,.
\end{equation}
Since in the presence of magnetic type interactions the optical geometry is no longer  Riemannian  \cite{Gibbons:2008zi},  we are led to replace (\ref{effact}) with the more general ansatz 
\begin{equation} \label{effds}
  {L_J \equiv \; } {\dot{s}_{\!J}} { \; = p_i\dot{x}^i\quad =
\quad F(\bm x,\dot{\bm x})}
\end{equation}
with $F$ an arbitrary homogeneous function of degree one in the second set of variables fulfilling appropriate regularity conditions. \smallskip
 
\noindent
From (\ref{maglag})  the canonical momenta and energy are given by:
\begin{equation} \label{magmom}
\begin{split}
p_i &= \frac{\partial L}{\partial \dot x^i} =  m{G}_{ij} (\bm x) \dot x^j + A_i (\bm x) \quad \Rightarrow \quad {mG}_{ij} (\bm x) \dot x^j = p_i - A_i (\bm x) = \pi_i \\
E &= p_i \dot x^i - L = \frac1{2m} {G}^{ij} (\bm x) \pi_i \pi_j + U (\bm x)
\end{split}
\end{equation}
Thus, by direct formulation by applying (\ref{magmom}) to 
(\ref{effds}), the Jacobi {Lagrangian} 
is given by:
\[ \begin{split}
{F(\bm x,\dot{\bm x})}
= p_i \dot x^i &= m G_{ij} (\bm x) \dot x^i \dot x^j + A_i (\bm x)
\dot x^i = \sqrt{2m \left( E - U (\bm x) \right) G_{ij} (\bm x) \dot x^i \dot x^j} + A_i (\bm x) \dot x^i 
\end{split} 
\]
Thus, the Jacobi metric for a system with magnetic fields is a
Finsler metric {of Randers type} given by:
{\begin{equation}\label{magjmet}
    ds_{\!J}=\sqrt{2m \left( E - U (\bm x)
  \right) G_{ij} (\bm x) dx^i dx^j} + A_i (\bm x) dx^i.
\end{equation}} 

This represents a generalization of the the original model of
Randers \cite{Randers:1941gge} which was used in
\cite{Gibbons:2008zi} to describe null geodesics in stationary spactimes.
\noindent We will compare this result to the non-relativistic limit of
the relativistic Jacobi metric that we will deduce later.

\section{Jacobi metric for a stationary spacetime}

Upon applying (\ref{mom}), the Jacobi Lagrangian is given by:
$$L_J = p_i \dot x^i = - mc \frac{{c} g_{0i} (\bm x) \dot t  \dot x^i + g_{ij} (\bm x) \dot x^i \dot x^j}{\sqrt{g_{\alpha \beta} (\bm x)  \dot x^\alpha \dot x^\beta}} \quad = \quad p_0 \frac{g_{0i} (\bm x)}{g_{00} (\bm x)} \dot x^i  + mc \frac{\left( {-}g_{ij} (\bm x) + \dfrac{g_{0i} (\bm x) g_{0j} (\bm x)}{g_{00} (\bm x)} \right) \dot x^i \dot x^j}{\sqrt{g_{\alpha \beta} (\bm x)  \dot x^\alpha \dot x^\beta}}$$
Here, let us define the spatial metric,
\begin{equation}
\gamma_{ij} (\bm x) := - g_{ij} (\bm x) + \frac{g_{0i} (\bm x) g_{0j} (\bm x)}{g_{00} (\bm x)},
\label{spatialmetric}
\end{equation}
allowing us to write the Jacobi Lagrangian as:
\begin{equation} \label{jlag} 
  L_J = p_0 \frac{g_{0i} (\bm x)}{g_{00} (\bm x)} \dot x^i + mc \sqrt{\frac{ \gamma_{ij} (\bm x)
  \dot x^i \dot x^j}{g_{\alpha \beta} (\bm x)  \dot x^\alpha \dot x^\beta}} \sqrt{ \gamma_{ij} (\bm x) \dot x^i \dot x^j}.
\end{equation}
From the 1st equation of (\ref{mom}) for momentum $p_0$, we obtain the following relationship,
\begin{equation} \label{condition}
p_0^2 = ( mc )^2 g_{00} (\bm x) \left( 1 + \frac{\gamma_{ij} (\bm x) \dot x^i \dot x^j}{g_{\alpha \beta} (\bm x)  \dot x^\alpha \dot x^\beta} \right) \qquad \Rightarrow \qquad \frac{\gamma_{ij} (\bm x) \dot x^i \dot x^j}{g_{\alpha \beta} (\bm x)  \dot x^\alpha \dot x^\beta} = \frac{{p_0^2-( mc )^2 g_{00} (\bm x)}}{( mc )^2 g_{00} (\bm x)}.
\end{equation}
Now, applying (\ref{condition}) to the Jacobi Lagrangian (\ref{jlag}), we have the form
\begin{equation} \label{jmet}
L_J = 
{F(\bm x,\dot{\bm x})} 
 = p_0 \frac{g_{0i} (\bm x)}{g_{00} (\bm x)} \dot x^i  
 + \sqrt{\frac{{p_0^2-( mc )^2 g_{00} (\bm x)}}{g_{00} (\bm x)} \gamma_{ij} (\bm x) \dot x^i \dot x^j},
\end{equation}
so that we can write the Jacobi metric as:

\begin{equation} \label{jacobimetric}
ds_{\!J}= \sqrt{\frac{p_0^2-(mc)^2 g_{00} (\bm x)}{g_{00} (\bm x)} \gamma_{ij} (\bm x) dx^i dx^j}+p_0 \frac{g_{0i} (\bm x)}{g_{00} (\bm x)} dx^i \equiv \sqrt{a_{ij}(\bm x)dx^i dx^j} +b_i dx^i\,.
\end{equation}
This result, which agrees with proposition 3.3 of \cite{Perlick},
is a Finsler metric of Randers type characterized by a Riemannian metric $a_{ij}$ and a one-form $b_i$ subject to positivity and convexity of $F$, which turn out to be satisfied provided that (e.g., \cite{bcs00}, p. 283f)
\begin{equation}
\sqrt{a^{ij}b_ib_j}<1\,.
\end{equation}
If we set $g_{0i} (\bm x) = 0$, the result (\ref{jacobimetric}) will clearly match (\ref{cnstrjm}). Under the approximation of non-relativistic limit, with weak potentials, we will obtain the non-relativistic limit of the Jacobi metric (\ref{jacobimetric}) as shown in \cite{jmp}:
$$g_{00} (\bm x) = 
1 + \frac{2 U{(\bm x)}}{mc^2} 
\quad , \quad - p_0 \; = \frac{\mathcal E}c \approx mc
+ \frac{E}{{c}} \quad , \quad E \ll mc^2 \quad , \quad 2 U (\bm x)  \ll  mc^2$$
$$\frac{p_0^2 - ( mc )^2 g_{00} (\bm x)}{g_{00} (\bm x)} \approx 2m \left( E - U (\bm x) \right) \qquad , \qquad  \frac{p_0}{g_{00} (\bm x)} \approx - mc \left( 1 + \frac{E - 2 U (\bm x)}{mc^2} \right) \approx - mc$$
$$g_{0i} (\bm x) g_{0j} (\bm x) \ll 1 \qquad \Rightarrow \qquad \gamma_{ij} (\bm x) \approx - g_{ij} (\bm x)$$

Thus, the non-relativistic Jacobi metric will be
\begin{equation} \label{nrjmet1}
{ds_{\!J}\approx\sqrt{-2m(E-U(\bm x))g_{ij}dx^idx^j} - m c g_{0i}(\bm x)dx^i}
\end{equation}
which is comparable to the Jacobi metric derived for a classical Lagrangian system with magnetic fields involved (\ref{magjmet}) if we write
$G_{ij}(\bm x)=-g_{ij}(\bm x)$ and $A_i(\bm x)=- m c g_{0i}(\bm x)$.

\section{Some examples of Jacobi metrics}

Two of the examples to which the formulation was applied require correction: the Taub-NUT, and Kerr metrics. The correct Jacobi-Maupertuis metrics are as follows:

\subsection{Taub-NUT metric}

The Euclidean Taub-NUT metric is given by:
\begin{equation} \label{taubnut}
d l^2 = 4 M^2 \frac{r - M}{r + M} \left( d \psi + \cos \theta \; d \varphi \right)^2 + \frac{r + M}{r - M} d r^2 + \left( r^2 - M^2 \right) \left( d \theta^2 + \sin^2 \theta \; d \varphi^2 \right) \, . 
\end{equation}
From which, we can deduce:
\begin{equation} \label{tnmet}
\begin{split}
g_{00} (\bm x) &= 4 M^2 \frac{r - M}{r + M} \qquad , \qquad g_{0 \varphi} (\bm x) = 4 M^2 \frac{r - M}{r + M} \ \cos \theta, \\ 
\gamma_{ij} (\bm x) \; &dx^i dx^j = - \frac{r + M}{r - M} d r^2 - \left( r^2 - M^2 \right) \left( d \theta^2 + \sin^2 \theta \; d \varphi^2 \right) \,.
\end{split}
\end{equation}
\noindent
Thus, according to (\ref{jacobimetric}),
\vskip .25 cm
for $p_\psi = - mc \dfrac{\partial \ }{\partial \dot \psi} \left( \sqrt{ \left(\dfrac{d l}{d \tau} \right)^2} \right)$ the Jacobi metric for Taub-NUT will be:
{\begin{equation} \label{tnjmet} 
\begin{split}
ds_{\!J}=&\sqrt{\left((mc)^2-\frac{p_\psi^2}{4 M^2} \frac{r + M}{r - M} \right)\left(\frac{r + M}{r - M} d r^2 + \left( r^2 - M^2 \right) \left( d \theta^2 + \sin^2 \theta \; d \varphi^2 \right)\right)}\\
& + p_{\psi} cos\theta \; d\varphi \,.
\end{split} 
\end{equation}
}

and in the low energy ($- p_\psi \approx 2M \left( mc - Q \right) \ , \ Q^2 \ll (mc)^2$) and weak field limit $M \ll r$, where
$$\frac{p_\psi^2}{4 M^2} \approx  (mc)^2 - 2 mc Q,$$
$$\frac{r + M}{r - M} \approx \left(1 + \frac Mr \right) \left(1 - \frac Mr \right) \approx 1 \quad , \quad \frac{p_\psi^2}{4 M^2} \left( \frac{r + M}{r - M} \right) - (mc)^2 \approx - 2mc Q$$
the Jacobi metric (\ref{tnjmet}) for Taub-NUT will become:
{\[
\begin{split}
ds_{\!J}=&\sqrt{2mcQ \left[ dr^2 + \left( r^2 - M^2 \right) \left( d \theta^2 + \sin^2 \theta \; d \varphi^2 \right) \right]}\\
&- 2M(mc-Q) \cos\theta \; d\varphi \,.
\end{split} \]
}

\subsection{Kerr metric}

The Kerr metric (for $c = 1$) is given by:
\begin{equation} \label{kerr}
\begin{split}
d l^2 &= \left( 1 - \frac{2 GM r}{\rho^2} \right) d t^2 + \frac{4 GM a r \sin^2 \theta}{\rho^2} d \varphi \; d t \\ 
& \qquad - \left[ \frac{\rho^2}{\Delta} dr^2 + \rho^2 d \theta^2 + \frac{\sin^2 \theta}{\rho^2} \left\{ \left( r^2 + a^2 \right)^2 - a^2 \Delta \sin^2 \theta \right\} d \varphi^2 \right] \,,\\ \\
\text{where } &\quad \Delta (r) = r^2 - 2 GM r + a^2 \qquad , \qquad \rho^2 (r, \theta) = r^2 + a^2 \cos^2 \theta \,.
\end{split}
\end{equation}
From which, we can deduce:
\begin{equation} \label{kerrmet}
\begin{split}
&g_{00} (\bm x) = 1 - \frac{2GM r}{\rho^2} \qquad, \qquad g_{0 \varphi} (\bm x) = \frac{2GM a r \sin^2 \theta}{\rho^2}, \\ 
&\gamma_{ij} (\bm x) \; dx^i dx^j = \rho^2\left(\frac{dr^2}{\Delta}+d\theta^2+
\frac{\Delta\sin^2\theta}{\Delta-a^2\sin^2\theta}d\varphi^2\right)
\end{split}
\end{equation}
Thus, according to (\ref{jacobimetric}), for $p_0=-\mathcal{E}$ the Jacobi metric for Kerr metric will be:
\begin{equation} \label{kerrjmet}
\begin{split}
ds_{\!J} = &\sqrt{\left(\frac{\mathcal{E}^2 \rho^2}{\Delta-a^2\sin^2\theta} - m^2\right)\rho^2\left(\frac{dr^2}{\Delta}+d\theta^2+
\frac{\Delta\sin^2\theta}{\Delta-a^2\sin^2\theta}d\varphi^2\right)}+
\\
&-\frac{2\mathcal{E}GMar\sin^2\theta}{\Delta-a^2\sin^2\theta}d\varphi.
\end{split}
\end{equation}
For $m=0$ and in the ultra-relativistic limit $\mathcal{E}\gg m$, this metric reduces to the Kerr-Randers optical metric investigated in \cite{Werner2012} up to a constant factor $\mathcal{E}$. For the non-relativistic low energy and weak potential limit ($GM r \ll \rho^2$)
$$U (\bm x) = - \frac{GMmr}{\rho^2} \quad , \quad |2 U (\bm x)| \ll m \quad \Rightarrow \quad \frac{2GMr}{\rho^2} \ll 1$$ 
$$\Rightarrow \quad \left( g_{0 \varphi} (\bm x) \right)^2 = \left( \frac{GMr}{\rho^2} \right)^2 4 a^2 \sin^4 \theta \ll 1$$
$$\Delta = \rho^2 \left( 1 - \frac{2 GM r}{\rho^2} \right) + a^2 \sin^2 \theta \approx \rho^2 + a^2 \sin^2 \theta \quad \Rightarrow \quad \Delta - a^2 \sin^2 \theta \approx \rho^2$$
$$- g_{ij} (\bm x) \; dx^i dx^j \approx \frac{\rho^2}{\rho^2 + a^2 \sin^2 \theta} dr^2 + \rho^2 d \theta^2 + \left( \rho^2 + a^2 \sin^2 \theta \right) \sin^2 \theta d \varphi^2$$
 we have according to (\ref{nrjmet1}):
\[\begin{split}
ds_{\!J}\approx&\sqrt{2 m \left(E + \frac{GMr}{\rho^2} \right)
\left( \frac{\rho^2}{\rho^2 + a^2 \sin^2 \theta} dr^2+\rho^2d\theta^2 + \left( \rho^2 + a^2 \sin^2 \theta \right) \sin^2 \theta d\varphi^2\right)}\\
&-m\frac{2GM a r \sin^2 \theta}{\rho^2}d\varphi. 
\end{split}\]

\section{{Relation to the gravitational magnetoelectric effect}}

The magnetoelectric effect refers to the property of certain materials (e.g., multiferroics) by which electric fields yield magnetization, and magnetic fields yields polarization. The linear magnetoelectric susceptibility $\alpha^{ij}= - \alpha^{ji}$ can be defined by the constitutive relations \footnote{in his section we use units in which the permeability and permittivity of the vacuum 
are set to unity and hence $c=1$} between the electromagnetic fields,
\begin{align}
D^i&=\varepsilon^{ij}E_j+\alpha^{ij}H_j\,, \\\
B^i&=\mu^{ij}H_j+\alpha^{ji}E_j\,,
\end{align}
where $\varepsilon^{ij}$ is the electric permittivity, and $\mu^{ij}$ the magnetic permeability. Now it turns out that electromagnetism in spacetime with $g_{0i}\neq0$ is also subject to this effect, which is thus called the gravitational magnetoelectric effect. But since this involves the {\it spatial} electromagnetic fields, its precise form depends both on the chart used and on the definition of the electromagnetic fields, for which two conventions are in common use. These involve tensor densities and may be referred to as the zero weight formalism and the unit weight formalism (for details, see e.g. \cite{GibbonsWerner}).

We adopt the latter which has the feature that in \emph{any} local coordinate system $(t,x^i)$ the source-free Maxwell's equations take the familiar form,\footnote{{The relation of the spatial electromagnetic fields ${\bf E}, {\bf B}, {\bf H}, {\bf D}$ to the components of the Maxwell tensor
 $F_{\mu \nu}$ is spelt out in detail in \S II.C of \cite{GibbonsWerner},
where quantities in the unit weight formalism are denoted by tildes, however.}}

\begin{eqnarray}
\nabla \times {\bf E} &=& - \frac{\partial  {\bf B}}{\partial t } \,,\qquad \nabla \cdot {\bf B}=0 \,,\\
\nabla \times  {\bf H} &=&  \frac{\partial {\bf D}}{\partial t }
\,,\qquad \nabla \cdot  {\bf D}=0 \,.
\end{eqnarray}

\begin{equation}
\gamma_{ij} = -g_{ij} + \frac{g_{0i} g_{0j}}{g_{00}}  \,, \qquad \gamma^{ij}= -g^{ij}  \,,
\end{equation}  
then the permittivity and permeability are found to be equal (a property known as {\it impedance-matched})
\begin{equation}
\varepsilon^{ij} = \mu^{ij} =  \frac{\sqrt{-g}}{g_{00}}\gamma^ {ij}= -\sqrt{-g} \frac{g^{ij} } {g_{00}}  \,, \label{perm}
\end{equation}
and the gravitational magnetoelectric susceptibility becomes
  \begin{equation}
  \alpha  ^{ij} =-\epsilon ^{ijk} \frac{g _{0k}}{ g_{00}} \,,
\label{alpha}
\end{equation}
where $\epsilon^{123}=\epsilon_{123}=1$, whence
\begin{equation}\begin{split} 
  g_{0k}= - \frac{1}{2}  g_{00}  \epsilon _{ijk} \alpha^{ij} \,.
\end{split}  
\end{equation}
Moreover, from (\ref{perm}) we have 
\begin{equation} \begin{split}
\gamma_{ij}=  \frac{g_{00}} {\sqrt{-g}} (\epsilon ^{-1})_{ij}
  = \frac{g_{00}}{\sqrt{-g} }  (\mu ^{-1})_{ij}\,.
\end{split}
\end{equation}
As before, we can now consider the geodesic of a free particle of mass $m$ in this spacetime. Then the Jacobi Lagrangian is the Finsler function of Randers form given by (\ref{jacobimetric})
and thus, using (\ref{perm}) and (\ref{alpha}), we find,
\begin{align}
  F(x,\dot x^i) &= p_0 \frac{ g_{0i}}{g_{00}}
\dot x^i +
 \sqrt{ \frac{p_0^2-m^2 g_{00}}{g_{00}} \gamma_{ij} \dot x^i \dot x^j   }\nonumber  \\
&= -\frac{1}{2}p_0 \epsilon _{ijk} \alpha ^{ij} \dot x ^k + \sqrt{
 \frac{ p_0^2 -    m^2 g_{00}}{\sqrt{-g} }
  (\epsilon ^{-1})_{ij} \dot x^i \dot x^j   } \,.
\end{align}

The inverse of the electric permittivity and the dual of the magnetoelectric susceptibility act therefore as the Riemannian metric and the one-form of the spatial Randers metric respectively. Of course by  (\ref{perm}) we obtain an identical relation in terms of the   magnetic  permeability and the dual of the magnetoelectric susceptibility.

\section{Conclusion}

In this paper we have studied the motion of massive particles
moving in a general stationary spacetime. We have obtained  
an explicit formula for the energy dependent Randers-Finsler metric
from which the equations of motion may be obtained. This extends and unifies
previous work  on the static case when  our result
reduces to a Jacobi-Maupertuis  style  Riemannian metric
and on the massless case  in stationary spacetimes,  in which case
our expression reduces to the previously known energy independent
Randers-Finsler metric. In the process we  have corrected some
misstatements in the literature.  We have given 
explicit formulae for our Jacobi-Maupertuis Randers-Finsler
metric in the astrophysically important  case of a Kerr black hole.
Finally we have studied Maxwell's equations in a general stationary metric
and shown how the effective electric permittivity , magnetic permeability
and magnetoelectric
susceptibilities  enter  our Jacobi-Maupertuis Randers-Finsler metric.
We believe that  this should contribute to our
understanding of polarization  due to gravitational lensing.


\begin{thebibliography}{99}

\bibitem{gwg} G.W. Gibbons, {\it The Jacobi-metric for timelike geodesics in static spacetimes}, \href{http://iopscience.iop.org/article/10.1088/0264-9381/33/2/025004/meta}{Class. Quantum Grav. {\bf 33} (2016) 025004}, arXiv: \href{http://arxiv.org/abs/1508.06755}{1508.06755}.

\bibitem{Gibbons:2008zi}
G.W. Gibbons, C.A.R. Herdeiro, C.M. Warnick and M.C. Werner,
\textit{Stationary Metrics and Optical Zermelo-Randers-Finsler Geometry}, Phys.\ Rev.\ D {\bf 79} (2009) 044022 doi:10.1103/PhysRevD.79.044022 [arXiv:0811.2877 [gr-qc]].
  
\bibitem{Randers:1941gge}
G.~Randers, \textit{On an Asymmetrical Metric in the Four-Space of General Relativity}, Phys.\ Rev.\  {\bf 59} (1941) no.2, 195. doi:10.1103/PhysRev.59.195

\bibitem{paolo} {G.W. Gibbons and} P. Maraner, {\em Comment on `The Jacobi metric for timelike geodesics in static spacetimes' (2016 Class. Quantum Grav. {\bf 33} 025004)}, preprint.

  
\bibitem{jmp} S. Chanda, G.W. Gibbons and P. Guha, {\em Jacobi-Maupertuis-Eisenhart metric and geodesic flows}, \href{http://aip.scitation.org/doi/full/10.1063/1.4978333}{J. Math. Phys. {\bf 58} (2017) 032503}, arXiv: \href{https://arxiv.org/abs/1612.00375}{1612.00375}.

\bibitem{FrolovShoom}  V. P. Frolov and A.A. Shoom, {\it Spinoptics in a stationary spacetime},  Phys. Rev. D {\bf 84}, 044026 (2011).

\bibitem{Castillo} G.F. Torres del Castillo and J. Mercado-P\'erez, {\it Three-dimensional formulation of the Maxwell  equations for stationary spacetimes}, J. Math. Phys. {\bf 40}, 2882-2890 (1999).

\bibitem{GibbonsWerner} G.W. Gibbons and M.C. Werner, {\it The gravitational magnetoelectric effect},  arXiv:1903.00223 [gr-qc].

\bibitem{Perlick}V.~Perlick, The brachistochrone problem in a
stationary space-time, J.\  Math,\  Phys.\  {\bf 32} (1991),
3148 ; https://doi.org/10.1063/1.529472.

\bibitem{cg} S. Chanda, and P. Guha, {\em Geometrical formulation of relativistic mechanics},
\href{https://www.worldscientific.com/doi/abs/10.1142/S0219887818500627}{Int. J. Geom. Meth. Mod. Phys. {\bf 15} (2018) 1850062},
arXiv: \href{https://arxiv.org/abs/1706.01921}{1706.01921}.




\bibitem{bcs00} D. Bao, S.-S. Chern and Z. Shen, {\it An Introduction to Riemann-Finsler Geometry} (Springer, New York, 2000).


\bibitem{Werner2012}
M. C. Werner, \textit{Gravitational lensing in the Kerr-Randers optical geometry}, Gen. Rel. Gravit. \textbf{44} (2012) 3047-3057. 



\end{thebibliography}
\end{document}